\documentstyle[twocolumn,aps,prd,epsf]{revtex}
\tightenlines

\def\d{{\rm d}}

\def\res#1#2{$\!\!#1\pm #2\!\!$}

\newcommand\BZ{\chi_{{\rm BZ}}}
\newcommand\NR{\theta^*_{{\rm NR}}}
\newcommand\KSW{\phi_{{\rm KSW}}}
\newcommand\jetalpha{\alpha_{{\rm 34}}}
\newcommand\as{{\alpha_s}}
\newcommand\NLO{next-to-leading order }
\newcommand\qb{{\bar q}}
\newcommand\Qb{{\bar Q}}
\newcommand{\Nc}{{N_C}}
\newcommand{\NA}{{N_A}}
\newcommand{\Nf}{{N_{\rm f}}}

\newcommand\Tr{{\rm Tr}}
\newcommand{\beq}{\begin{equation}}
\newcommand{\eeq}{\end{equation}}
\newcommand{\beqn}{\begin{eqnarray}}
\newcommand{\eeqn}{\end{eqnarray}}

\newcommand\ycut{y_{\rm cut}}

\begin{document}

\onecolumn

\preprint{hep-ph/9712385}
\title{Four-jet angular distributions and color charge measurements:
leading order versus next-to-leading order}
\author{Zolt\'an Nagy$^a$ and Zolt\'an Tr\'ocs\'anyi$^{b,a}$}
\address{$^a$Department of Theoretical Physics, KLTE,
H-4010 Debrecen P.O.Box 5, Hungary
\\
$^b$Institute of Nuclear Research of the Hungarian Academy of Sciences,
H-4001 Debrecen P.O.Box 51, Hungary}
\date{\today}
\maketitle

\begin{abstract}
We present the next-to-leading order perturbative QCD prediction to the
four-jet angular distributions used by experimental collaborations at
LEP for measuring the QCD color charge factors. We compare our results
to ALEPH data corrected to parton level.  We perform a leading order
``measurement'' of the QCD color factor ratios by fitting the leading
order perturbative predictions to the next-to-leading order result.
Our result shows that in an experimental analysis for measuring the
color charge factors the use of the O($\alpha_s^3$) QCD predictions
instead of the O($\alpha_s^2$) results may shift the center of the fit
by a relative factor of $1+2\as$ in the $T_R/C_F$ direction.
\end{abstract}

\section{Introduction}

In the first phase of operation of the Large Electron Positron
Collider (LEP) four-jet events were primarily used for measuring the
eigenvalues of the Casimir operators of the underlying symmetry group,
the QCD color factors \cite{L3,a34,DELPHI,OPAL,ALEPH}. The values of
these color charges test whether the dynamics is indeed described by an
SU(3) symmetry. The dependence of jet cross sections on the adjoint
color charge appear at O($\alpha_s^2$). Several test variables with
perturbative expansion starting at O($\alpha_s^2$) --- so called four-jet
angular distributions --- were proposed as candidate observables with
particular sensitivity to the gauge structure of the theory
\cite{a34,KSW,NR,BZ}.  For a long time, however, perturbative QCD
prediction for these variables at O($\alpha_s^3$) had not been
available, therefore the absolute normalization of the perturbative
prediction could not be fixed. In order to circumvent this problem the
experimental collaborations either fitted the strong coupling as well,
or used normalized angular distributions of four jet events that were
expected to be insensitive to renormalization scale dependence. The
small scale dependence however, is but an indication and not a proof of
negligible radiative corrections: in principle the shape of the
distribution can change from O($\alpha_s^2$) to O($\alpha_s^3$).
Therefore, it is desirable to check explicitly the effect of the \NLO
corrections on these normalized angle distributions.

Recently the \NLO corrections to various four-jet observables have been
calculated \cite{DSjets,Signer,NT}. These works depend crucially on the
matrix elements for the relevant QCD subprocesses, i.e.\ for the
$e^+e^-\to \qb q gg$ and $e^+e^-\to \qb q \Qb Q$ processes at one
loop and for the $e^+e^-\to \qb q ggg$ and $e^+e^-\to \qb q \Qb Q g$
processes at tree level. The loop results became available due to the
effort of two groups. In refs.~\cite{GM4q,CGM2q2g} Campbell, Glover
and Miller made FORTRAN programs for the \NLO squared matrix elements
of the $e^+e^- \to \gamma^* \to \qb q \Qb Q$ and $\qb q g g$ processes
publicly available. In refs.~\cite{BDKW4q,BDK2q2g} Bern, Dixon, Kosower
and Wienzierl gave analytic formul\ae\ for the helicity amplitudes of
the same processes with the $e^+e^- \to Z^0 \to$ four partons channel
included as well. The helicity amplitudes for the five-parton processes
have been known for a long time \cite{a5parton}. Using the helicity
amplitudes in refs.~\cite{BDKW4q,BDK2q2g,a5parton}, Dixon and Signer
calculated the \NLO corrections to four-jet fractions for various
clustering algorithms \cite{DSjets}, as well as to the $\chi_{\rm BZ}$
angle distribution \cite{Signer}. In previous publications we calculated
several four-jet shape variables at the \NLO accuracy \cite{NT}. In this
paper we calculate the radiative corrections to the distributions of
the commonly used angular shape variables $\KSW$, $\NR$, $\jetalpha$, and
repeat the calculation for the distribution of $\BZ$.  We use the
matrix elements of refs.~\cite{BDKW4q,BDK2q2g} for the loop
corrections, while calculated the helicity amplitudes of the relevant
tree-level processes ourselves.

Knowing the \NLO corrections to these angular distributions, one would
like to quantify their effect on the measurement of the QCD color charges.
We estimate the systematic error coming from the use of leading order
results instead of the \NLO one in the fits for the color charge
ratios $x = C_A/C_F$ and $y= T_R/C_F$ in the following way. We assume
that \NLO QCD is the true theory that describes the data. We fit the
leading order prediction of the angular distributions with $x$ and $y$ 
left free to our \NLO QCD results and determine these charge ratios from
this fit. The central value of the ``measured'' charge ratios differs
from the SU(3) values $x = 9/4$ and $y = 3/8$. This shift is the
systematic bias that comes from the use of leading order fits in
experimental analyses instead of the \NLO ones.

The rest of the paper is organized as follows. In section II we outline
the structure of the numerical calculation and describe how we
parametrise our results. In section III we present the complete
O($\alpha_s^3$) predictions for the four standard angular distributions
using two different jet algorithms: the Durham algorithm \cite{durham}
and the Cambridge algorithm proposed recently \cite{cambridge}. In
section IV we perform the leading order fit of the color charges to our
\NLO results. Section V contains our conclusions.

\section{The structure of the numerical calculation}

It is well known that the \NLO correction is a sum of two integrals ---
the real and virtual corrections --- that are separately divergent (in
the infrared) in $d=4$ dimensions. For infrared safe observables, for
instance the four-jet angular distributions used in this work, their
sum is finite. In order to obtain this finite correction, we use a
slightly modified version of the dipole method of Catani and Seymour
\cite{CSdipole} that exposes the cancellation of the infrared
singularities directly at the integrand level. The formal result of
this cancellation is that the \NLO correction is a sum of two finite
integrals,
\beq                         
\label{sNLO}
\sigma^{\rm NLO} 
= \int_5 \d\sigma_5^{\rm NLO} + \int_4 \d\sigma_4^{\rm NLO}\:,
\eeq
where the first term is an integral over the available five-parton
phase space (as defined by the jet observable) and the second one is
an integral over the available four-parton phase space.

Once the phase space integrations in eq.~(\ref{sNLO}) are carried out,
the \NLO differential cross section for the four-jet observable $O_4$
takes the general form
%\beqn
%\label{nloxsec}
%&&\frac{1}{\sigma_0}\frac{\d \sigma}{\d O_4}(O_4)
%= \left(\frac{\as(\mu) C_F}{2\pi}\right)^2 B_{O_4}(O_4)
%\\ \nn &&\qquad
%+ \left(\frac{\as(\mu) C_F}{2\pi}\right)^3
%\left[B_{O_4}(O_4) \frac{\beta_0}{C_F}
% \ln\frac{\mu^2}{s} + C_{O_4}(O_4)\right]\:.
%\eeqn
\beq
\label{nloxsec}
\frac{1}{\sigma_0}\frac{\d \sigma}{\d O_4}(O_4)
= \left(\frac{\as(\mu) C_F}{2\pi}\right)^2 B_{O_4}(O_4)
+ \left(\frac{\as(\mu) C_F}{2\pi}\right)^3
\left[B_{O_4}(O_4) \frac{\beta_0}{C_F}
 \ln\frac{\mu^2}{s} + C_{O_4}(O_4)\right]\:.
\eeq
In this equation $\sigma_0$ denotes the Born cross section for the
process $e^+e^-\to \qb q$, $s$ is the total c.m.\ energy squared, $\mu$
is the renormalization scale, while $B_{O_4}$ and $C_{O_4}$ are
scale independent functions, $B_{O_4}$ is the Born approximation
and $C_{O_4}$ is the radiative correction. We use the two-loop
expression for the running coupling,
\beq
\as(\mu) = \frac{\as(M_Z)}{w(\mu)}
\left(
1-\frac{\beta_1}{\beta_0}\frac{\as(M_Z)}{2\pi}\frac{\ln(w(\mu))}{w(\mu)}
\right)\:,
\eeq
with
\beq
w(\mu) = 1 - \beta_0
\frac{\as(M_Z)}{2\pi}\ln\left(\frac{M_Z}{\mu}\right)\:,
\eeq
\beq
\beta_0 = \frac{11}{3}C_A - \frac{4}{3} T_R \Nf\:,
\eeq
\beq
\beta_1 = \frac{17}{3}C_A^2 - 2 C_F T_R \Nf - \frac{10}{3} C_A T_R \Nf \:,
\eeq
with the normalization $T_R=1/2$ in $\Tr(T^aT^{\dag b})=T_R\delta^{ab}$.
The numerical values presented in this letter were obtained at the $Z^0$
peak with $M_Z=91.187$\,GeV, $\Gamma_Z=2.49$\,GeV, $\sin_W^2\theta=0.23$,
$\as(M_Z)=0.118$ and $\Nf=5$ light quark flavors.

The Born approximation and the higher order correction are linear and
quadratic forms of ratios of the color charges \cite{NTloopamps}:
\beq                                                                  
B_4 = B_0 + B_x\,x + B_y\,y \:,
\eeq
and            
%\beqn
%\label{C4}
%&&C_4 = C_0 +\,C_x\,x + C_y\,y + C_z\,z              
%\\ \nn && \quad\;\;
%+\,C_{xx}\,x^2 + C_{xy}\,x\,y + C_{yy}\,y^2 \:.
%\eeqn
\beq
\label{C4}
C_4 = C_0 +\,C_x\,x + C_y\,y + C_z\,z              
+\,C_{xx}\,x^2 + C_{xy}\,x\,y + C_{yy}\,y^2 \:.
\eeq
At \NLO the ratio $z$ appears that is related to the square of a cubic
Casimir,
\beq
C_3 = \sum_{a,b,c=1}^\NA \Tr(T^a T^b T^{\dag c}) \Tr(T^{\dag c} T^b
T^a)\:,    
\eeq
via $z=\frac{C_3}{\Nc C_F^3}$. The Born functions $B_i$ are obtained by
integrating the fully exclusive O($\alpha_s^2$) ERT matrix elements
\cite{ERT} and were used by the experimental collaborations
\cite{L3,a34,DELPHI,OPAL,ALEPH}. In the next section we present the
$C_i$ correction functions for the four different angular distributions.

\section{Results}

In order to define the angular variables we denote the three-momenta of
the four jets by $\vec{p}_i$, ($i=1,2,3,4$) and label jets in order
of descending jet energy, such that jet 1 has the highest energy and
jet 4 has the smallest. The four variables are
\begin{enumerate}
\item
the K\"orner-Schierholz-Willrodt variable \cite{KSW},%\\
$\cos\KSW$ is the
cosine of the average of two angles between planes spanned by the jets,
%\beqn
%&&\KSW = \frac{1}{2}
%\left(\arccos
%\left(\frac{(\vec{p}_1\times\vec{p}_4)\cdot (\vec{p}_2\times\vec{p}_3)}
%{|\vec{p}_1\times\vec{p}_4| |\vec{p}_2\times\vec{p}_3|}\right)\right.
%\\ \nn &&\qquad\qquad\quad
%\left. + \arccos
%\left(\frac{(\vec{p}_1\times\vec{p}_3)\cdot (\vec{p}_2\times\vec{p}_4)}
%{|\vec{p}_1\times\vec{p}_3| |\vec{p}_2\times\vec{p}_4|}\right)\right)\:;
%\eeqn
\beq
\KSW = \frac{1}{2}
\left[\arccos
\left(\frac{(\vec{p}_1\times\vec{p}_4)\cdot (\vec{p}_2\times\vec{p}_3)}
{|\vec{p}_1\times\vec{p}_4| |\vec{p}_2\times\vec{p}_3|}\right)
+ \arccos
\left(\frac{(\vec{p}_1\times\vec{p}_3)\cdot (\vec{p}_2\times\vec{p}_4)}
{|\vec{p}_1\times\vec{p}_3| |\vec{p}_2\times\vec{p}_4|}\right)\right]\:;
\eeq
\item
the modified Nachtmann-Reiter variable \cite{NR}, $|\cos\NR|$ is the
absolute value of the cosine of the angle between the vectors
$\vec{p}_1-\vec{p}_2$ and $\vec{p}_3-\vec{p}_4$,
\beq
\cos\NR =
\frac{(\vec{p}_1-\vec{p}_2)\cdot (\vec{p}_3-\vec{p}_4)}
{|\vec{p}_1-\vec{p}_2| |\vec{p}_3-\vec{p}_4|}\:;
\eeq
\item
$\cos\jetalpha$ \cite{a34}, the cosine of the angle between the two
smallest energy jets,
\beq
\cos\jetalpha =
\frac{\vec{p}_3\cdot \vec{p}_4}{|\vec{p}_3| |\vec{p}_4|}\:;
\eeq
\item
the Bengtsson-Zerwas correlation \cite{BZ}, $|\cos\BZ|$ is the
absolute value of the cosine of the angle between the plane spanned by
jets 1 and 2 and that by jets 3 and 4,
\beq
\cos\BZ =
\frac{(\vec{p}_1\times\vec{p}_2)\cdot (\vec{p}_3\times\vec{p}_4)}
{|\vec{p}_1\times\vec{p}_2| |\vec{p}_3\times\vec{p}_4|}\:;
\eeq
\end{enumerate}
We tabulate the numerical value of the \NLO kinematic functions for
these angular variables in the Appendix in Tables III--VI for the Durham
clustering algorithm and in Tables VII--X for the Cambridge algorithm
that was proposed recently \cite{cambridge}.  Using this new algorithm,
the hadronization corrections are expected to be much smaller
therefore, the perturbative prediction is more reliable. The values in
the tables were obtained by selecting four-jet events at a fixed jet
resolution parameter $\ycut=0.008$ which is the value used by the ALEPH
collaboration \cite{ALEPH}. We do not show the value of the $C_z$
functions because they turn out to be negligible. The $C_4$ values were
obtained according to eq.~(\ref{C4}) with SU(3) values for the color
charge ratios, $x = 9/4$, $y = 3/8$. Comparing the size of
the corrections for these two algorithms, we see that in general the
$C_i$ functions in the case of the Cambridge algorithm are
10--20\,\% smaller.

We use the numerical values for the kinematic functions to calculate
the \NLO QCD predictions for the SU(3) values $x = 9/4$, $y = 3/8$
according to eq.~(\ref{nloxsec}) at $x_\mu=\mu/\sqrt{s}=1$. We compare
our predictions for the Durham algorithm (solid histograms) to ALEPH
data (diamonds) in Figs.~1--4. In order to make this
comparison we normalize the histograms to one, therefore
\beq
F(z) = \frac{1}{\sigma}\frac{\d \sigma}{\d z}(z)
\eeq
in the plots. The qualitative agreement between data and theory is
very good.  Also shown in these figures our results for the Cambridge
algorithm (dotted histograms). The statistical error of the Monte
Carlo integrals is below 1.5\,\% for the Durham
algorithm and below 2\,\% in the case of the Cambridge algorithm in
each of the bins. In the same figures, the windows show polinomial
fits to the $K$ factors of the normalized distributions defined as
\beq
K(z) = \frac{1}{\sigma_{{\rm NLO}}}\frac{\d\sigma_{{\rm NLO}}}{\d z}(z)
\Bigg/ \frac{1}{\sigma^{{\rm LO}}}\frac{\d\sigma^{{\rm LO}}}{\d z}(z)\:,
\eeq
where $\sigma_{\rm NLO}=\sigma^{\rm LO}+\sigma^{\rm NLO}$
is the \NLO cross section. The $\chi^2/N_{\rm dof}$ of these fits is between
(1.5--7)/20. The $K$ factors for the $|\cos\BZ|$ distributions, for the
$\cos\jetalpha$ distribution with Durham algorithm and for the $|\cos\NR|$
distribution with the Cambridge algorithm are approximately constant 1 
over the whole range, therefore the shape of the leading and \NLO 
distributions are very similar in these cases.
\begin{figure}
\epsfxsize=15cm \epsfbox{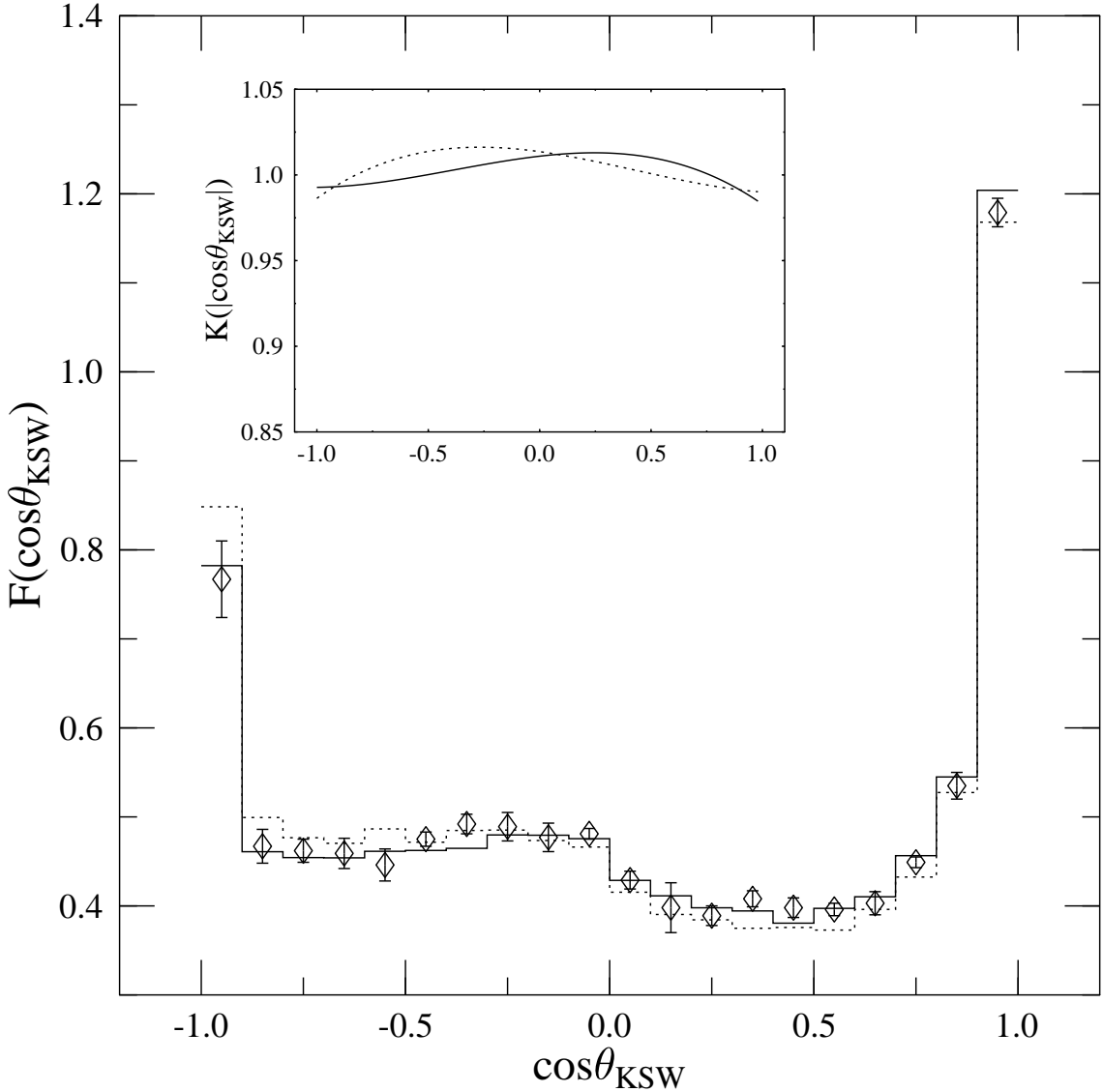}
\caption{Comparison of the \NLO QCD prediction for the $\cos\KSW$
distribution obtained using Durham (solid) and Cambridge (dotted)
jet algorithm with ALEPH data (diamonds). In the window the $K$ factor
of the distribution with Durham (solid) and with the Cambridge (dotted)
algorithm is shown.}
\end{figure}
\begin{figure}
\epsfxsize=15cm \epsfbox{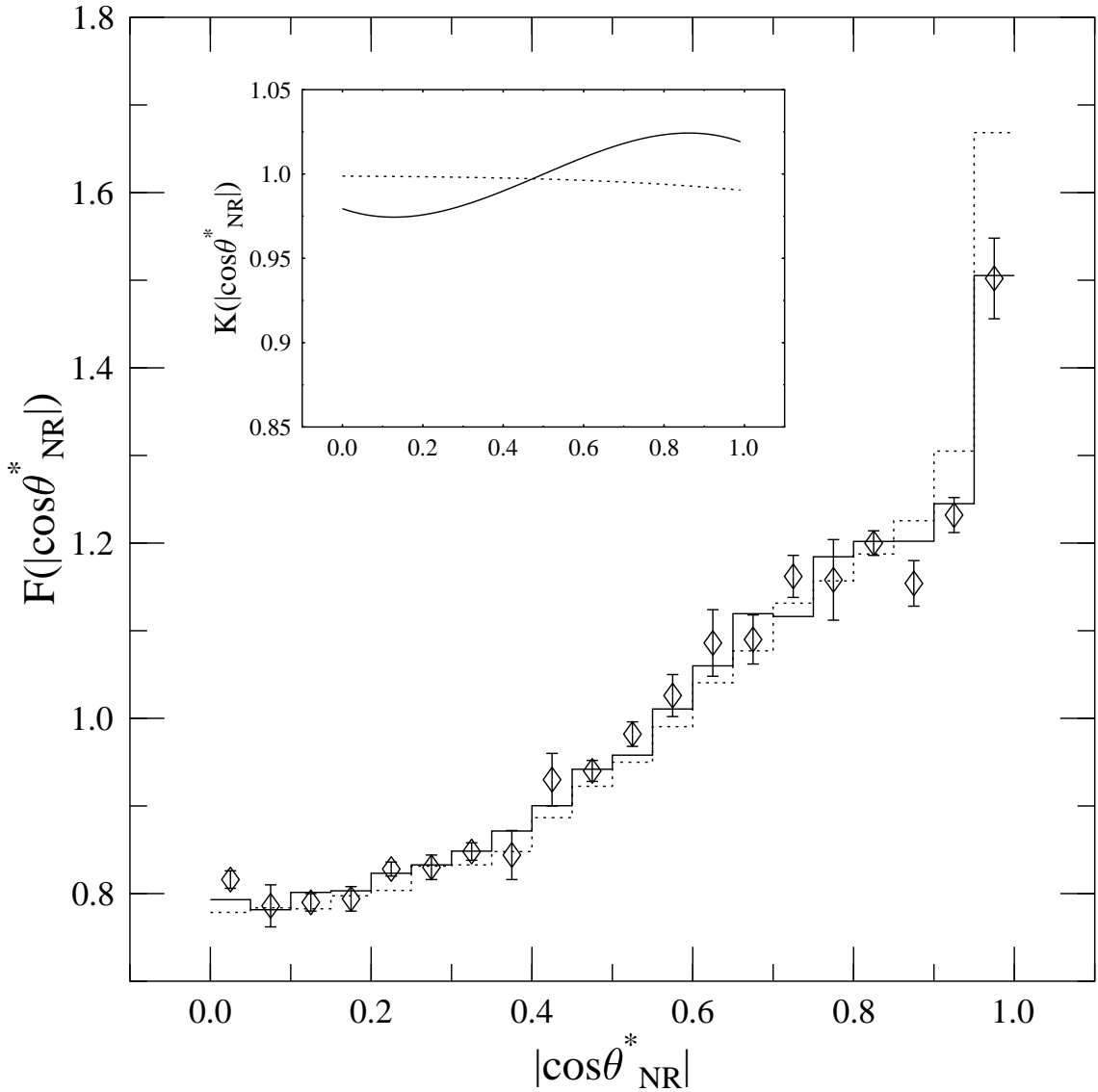}
\caption{Comparison of the \NLO QCD prediction for the $|\cos\NR|$
distribution obtained using Durham (solid) and Cambridge (dotted)
jet algorithm with ALEPH data (diamonds). In the window the $K$ factor
of the distribution with Durham (solid) and with the Cambridge (dotted)
algorithm is shown.}
\end{figure}
\begin{figure}
\epsfxsize=15cm \epsfbox{fig2.epsi}
\caption{Comparison of the \NLO QCD prediction for the $\cos\jetalpha$
distribution obtained using Durham (solid) and Cambridge (dotted)
jet algorithm with ALEPH data (diamonds). In the window the $K$ factor
of the distribution with Durham (solid) and with the Cambridge (dotted)
algorithm is shown.}
\end{figure}
\begin{figure}
\epsfxsize=15cm \epsfbox{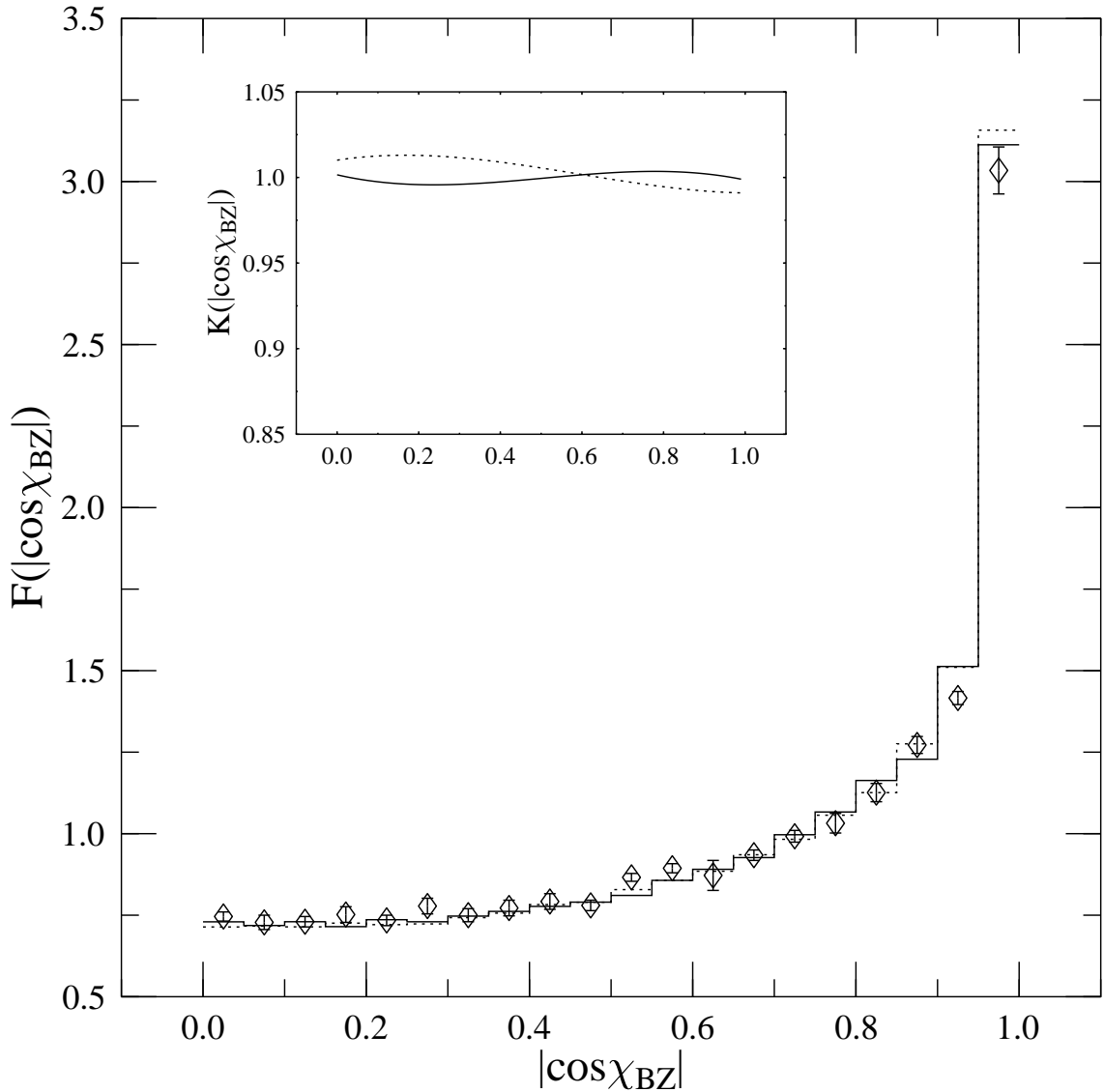}
\caption{Comparison of the \NLO QCD prediction for the $|\cos\BZ|$
distribution obtained using Durham (solid) and Cambridge (dotted)
jet algorithm with ALEPH data (diamonds). In the window the $K$ factor
of the distribution with Durham (solid) and with the Cambridge (dotted)
algorithm is shown.}
\end{figure}

\section{Leading order versus next-to-leading order}

A quantitative comparison of the data for the angular distributions
to the \NLO prediction decomposed in a quadratic form of the color
factor ratios with group independent kinematical functions as
coefficients makes possible a simultaneous fit of the strong coupling
and the color charge ratios.  That procedure would require a full
experimental analysis which is not our goal in the present paper. What
we would like to achieve is to give a reliable estimate of the
systematic theoretical uncertainty coming from the use of the leading
order perturbative prediction instead of the \NLO one in a color charge
measurement. To this end we pretend that the \NLO perturbative QCD is
the ``true'' theory that describes data perfectly. We ``produce'' data
using our \NLO prediction with SU(3) values $x=9/4$ and $y=3/8$ and
perform a leading order fit of $x$ and $y$ using the $B_0$, $B_x$ and
$B_y$ functions.  We use $\chi^2$ minimalization to obtain the best
values with
%\beqn
%&&\chi^2 = \\ \nn
%&&\sum_i \frac{1}{w_i^2}\left(
%\frac{B_0(z_i)+x B_x(z_i)+y B_y(z_i)}
%     {\sigma_0+x \sigma_x+y \sigma_y}
%-\frac{1}{\sigma_{\rm NLO}}\frac{\d\sigma_{\rm NLO}}{\d z}(z_i)\right)^2
%\eeqn
\beq
\chi^2 = 
\sum_i \frac{1}{w_i^2}\left(
\frac{B_0(z_i)+x B_x(z_i)+y B_y(z_i)}
     {\sigma_0+x \sigma_x+y \sigma_y}
-\frac{1}{\sigma_{\rm NLO}}\frac{\d\sigma_{\rm NLO}}{\d z}(z_i)\right)^2
\eeq
where $w_i$ is the statistical error of the normalized \NLO distribution
in the $i$th bin, the summation runs over the bins and
$\sigma_j(z_i)$ ($j = 0$, $x$, $y$, or NLO) defined as
\beq
\sigma_j=\int\!\d z\,B_j(z)\:,\quad
\sigma_{\rm NLO}=\int\!\d z\,\frac{\d\sigma_{\rm NLO}}{\d z}(z)
\eeq
As a check of the fit we also performed a linear fit to the not normalized
distributions in the form
%\beqn
%&&\chi^2 = \\ \nn
%&&\sum_i \frac{1}{w_i^2}\left( \eta (B_0(z_i)+x B_x(z_i)+y B_y(z_i))
%-\frac{\d\sigma_{\rm NLO}}{\d z}(z_i)\right)^2\:,
%\eeqn
\beq
\chi^2 = 
\sum_i \frac{1}{w_i^2}\left( \eta (B_0(z_i)+x B_x(z_i)+y B_y(z_i))
-\frac{\d\sigma_{\rm NLO}}{\d z}(z_i)\right)^2\:,
\eeq
where $w_i$ is the statistical error of the \NLO distribution in the
$i$th bin, and with $\eta = (\as C_F/(2\pi))^2$ fitted as well. The two
procedures give the same result for $x$ and $y$ to very good accuracy.

We performed the fit for each angular distribution separately, as well
as for the four angular variables combined. Table I contains the results
of these fits.
We see that the shifts in the $x$--$y$ values are quite large. Looking at
the errors, one finds that the shift is significant only if the $K$ factor
of the corresponding distribution (see Figs.~1--4) is not constant 1.
For those cases when the shapes of the leading order and the \NLO
distributions are very similar, i.e.\ the $K\simeq 1$, then the fits
give values compatible with the canonical QCD values. The origin of the
large errors in some fits is the global correlation between the two
parameters $x$ and $y$ in the fit. In these cases --- $|\cos\NR|$,
$\cos\jetalpha$ and $|\cos\BZ|$ distributions --- one cannot fit 
both variables reliably. Instead, one can fit either the ratio of
the two parameters, or fix one parameter to the SU(3) value and
fix the other. For instance, fixing $x = 9/4$ one obtains the fitted
values for $y$ as given in Table~II. We observe from Figs.~1--4 and
Table~II that in those cases, when $K\simeq 1$ the result of the fit is
in agreement with SU(3) --- Durham $\cos\jetalpha$, $|\cos\BZ|$ and 
Cambridge $|\cos\NR|$ distributions ---, while for the rest of the
distributions we obtain fit parameter different from the SU(3) value
because the shapes of the leading and \NLO distributions are different.
\vbox{
\begin{table}             
\caption{Leading order fit of the color charge ratios to the
next-to-leading order differential distributions of the angular correlations.}
\begin{tabular}{ccc}
  Observable     &       $x$       &       $y$    \\   
\tableline
                 &Durham algorithm &   \\
\tableline
 $\cos\KSW$      & $2.21 \pm 0.05$ & $0.58 \pm 0.07 $   \\
 $|\cos\NR|$     & $1.41 \pm 1.43$ & $0.08 \pm 0.11 $   \\
 $\cos\jetalpha$ & $2.08 \pm 0.21$ & $0.57 \pm 0.23 $   \\
 $|\cos\BZ|$     & $1.15 \pm 1.43$ & $0.12 \pm 0.31 $   \\
%A34,BZ          & $2.27 \pm 0.05$ & $0.36 \pm 0.04 $   \\
%BZ,NR           & $0.83 \pm 0.33$ & $0.04 \pm 0.05 $   \\
%A34,KSW         & $2.17 \pm 0.05$ & $0.53 \pm 0.05 $   \\
 all four        & $2.32 \pm 0.03$ & $0.29 \pm 0.02 $   \\
\tableline
                 &Cambridge algorithm &   \\
\tableline
 $\cos\KSW$      & $2.30 \pm 0.08$ & $0.52 \pm 0.09 $   \\
 $|\cos\NR|$     & $0.99 \pm 2.70$ & $0.21 \pm 0.31 $   \\
 $\cos\jetalpha$ & $0.34 \pm 0.42$ & $2.65 \pm 0.48 $   \\
 $|\cos\BZ|$     & $3.53 \pm 2.80$ & $0.82 \pm 0.68 $   \\
%A34,BZ          & $2.18 \pm 0.08$ & $0.55 \pm 0.05 $   \\
%BZ,NR           & $1.19 \pm 0.53$ & $0.25 \pm 0.09 $   \\
%A34,KSW         & $2.20 \pm 0.07$ & $0.60 \pm 0.08 $   \\
 all four        & $2.29 \pm 0.05$ & $0.45 \pm 0.03 $   \\
\end{tabular}
\end{table}
\noindent

\begin{table}             
\caption{Leading order fit the color charge ratio $y$ to the
next-to-leading order differential distributions of the angular
correlations with $x = 9/4$ fixed.}
\begin{tabular}{ccc}
  Observable     & Durham algorithm&Cambridge algorithm \\
\tableline
 $\cos\KSW$      & $0.57 \pm 0.06$ & $0.55 \pm 0.08 $   \\
 $|\cos\NR|$     & $0.15 \pm 0.03$ & $0.36 \pm 0.05 $   \\
 $\cos\jetalpha$ & $0.39 \pm 0.05$ & $0.56 \pm 0.08 $   \\
 $|\cos\BZ|$     & $0.35 \pm 0.05$ & $0.51 \pm 0.06 $   \\
%A34,BZ          & $0.37 \pm 0.03$ & $0.53 \pm 0.05 $   \\
%BZ,NR           & $0.22 \pm 0.03$ & $0.41 \pm 0.04 $   \\
%A34,KSW         & $0.47 \pm 0.04$ & $0.56 \pm 0.06 $   \\
 all four        & $0.31 \pm 0.02$ & $0.46 \pm 0.03 $   \\
\end{tabular}
\end{table}
\noindent

}

%\indent
We show the result of the combined fit of all four variables in Fig.~5
in the form of 68.3\,\% and 95\,\% confidence level contours in the
$x$--$y$ plane
with ellipses centered on the best $x$--$y$ pair. There are five
contours sitting on three different centers in each plot. The fits with
both 1- and 2-$\sigma$ contours were obtained using all four angular
distributions with all bins included. The fit with only 1-$\sigma$ contour
shown corresponds to the ``ALEPH choice'': using all four variables with 
fit ranges $0.1 \le |\cos\BZ|,\;|\cos\NR| \le 0.9$ and
$-0.8 \le \cos\jetalpha,\;\cos\KSW \le 0.8$.
%}

We observe from Fig.~5., that the leading order fit results in
overestimating the $C_A/C_F$ ratio by 2--3\,\% no matter which
clustering algorithm is used. For the $T_R/C_F$ ratio the leading order
fit underestimates the result by 20--30\,\% of a \NLO fit when the the
Durham algorithm is used, while in the case of Cambridge clustering the
leading order fit gives an overestimate of about 20\,\%. This systematic
bias appears significant in both cases. Although the two parameters are
slightly correlated when all four variables are used, the fit is reliable.
The result of the fit depends on the jet algorithm because the different
jet finders lead to different jet momenta from which our test variables
are built.
We also see that constraining the fit range as the ALEPH collaboration
did does not alter our conclusions significantly. We would like to
emphasize that the significant shift from the SU(3) values does not mean
the exclusion of QCD, but simply gives an estimate of the systematic 
theoretical error in the color charge measurements when leading order
fits are used.
\begin{figure}
\epsfxsize=15cm \epsfbox{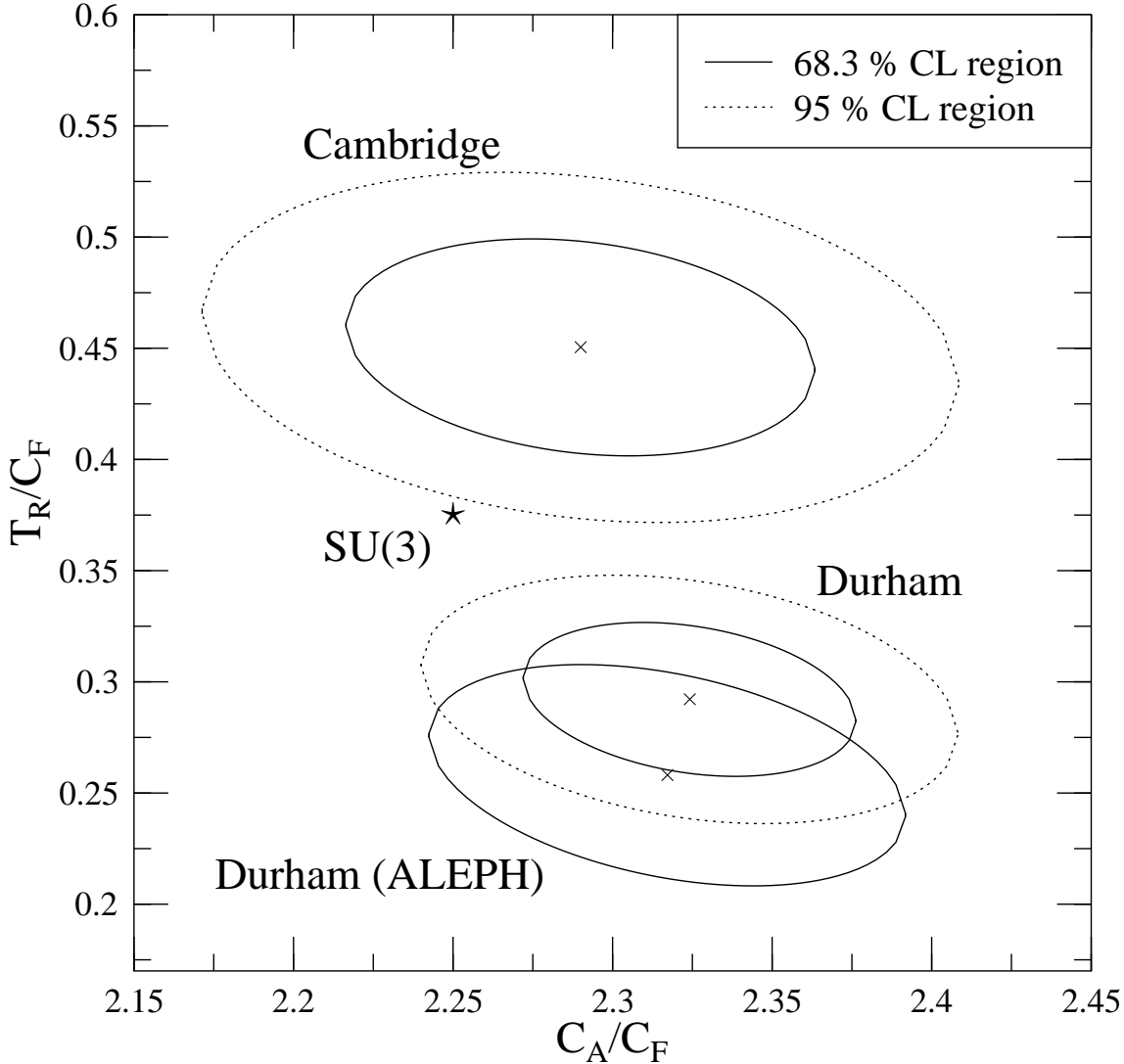}
\caption{Confidence level contours of the leading order fits of the color
charges $x$ and $y$ for the Durham and Cambridge clustering at $\ycut=0.008$.}
\end{figure}

One may ask how the light gluino exclusion significance changes in the
recent analysis of Csikor and Fodor \cite{Fodor}, which used the results
of four-jet analyses, if one takes into account the systematic theoretical
error discussed above. Assuming that the shifts of $x$ and $y$ are
similar in the light gluino extension of QCD, our conclusion suggests
that the radiative corrections induce a shift of order $2\alpha_s$ times
the tree-level value for $x$ and $y$. Lacking this piece of information
Csikor and Fodor have  increased the axes of the error ellipses  by a
factor of $\as$ times the theoretical $x$ and $y$ values. Implementing
our results  to a Csikor-Fodor type analysis for the four-jet events
would decrease their confidence levels for the light gluino exclusion
from 99.9\,\% (Csikor-Fodor value) to $\simeq$98\,\%, which is, however,
still much higher than a 2-$\sigma$ exclusion.

\section{Conclusion}

In this paper we presented the \NLO corrections to the group independent
kinematical functions of the four standard four-jet angular
distributions, $\cos\KSW$, $|\cos\NR|$, $\cos\jetalpha$, and
$|\cos\BZ|$ with jets defined with two --- the Durham and the Cambridge
--- clustering algorithms at $\ycut = 0.008$. These results were obtained
with a general purpose Monte Carlo program called DEBRECEN
\cite{debrecen} that can be used to calculate the
differential distribution of any other four-jet quantity at the \NLO
accuracy in electron-positron annihilation. The $|\cos\BZ|$ distribution
 using the Durham clustering algorithm was first calculated by Signer
\cite{Signer}. Our result agrees with his within statistical errors.
The results for the other three distributions with Durham
clustering as well as for all distributions when the Cambridge
algorithm is used are new. We compared our results to data obtained by
the ALEPH collaboration corrected to parton level and found very good
qualitative agreement. We have also presented the $K$ factors of the 
distributions using both jet clustering algorithms.

Having the \NLO perturbative QCD prediction at our disposal we made
an estimate of the systematic theoretical error of the QCD color charge
measurements due to the use of leading order group independent
kinematical functions. We found that the use of the O($\alpha_s^3$) QCD
predictions instead of the O($\alpha_s^2$) results may shift the center
of the fit by a relative factor of about $1+2\as$ in the
$T_R/C_F$ direction, while the best $C_A/C_F$ value is hardly affected.

\bigskip
%\large{\bf Acknowledgment: }\normalsize
We are grateful to F. Csikor, Z.  Fodor and S. Catani for useful
communication, and to G. Dissertori for providing us the ALEPH data.
This research was supported in part by the EEC Programme "Human Capital
and Mobility", Network "Physics at High Energy Colliders", contract
PECO ERBCIPDCT 94 0613 as well as by the Hungarian Scientific Research
Fund grant OTKA T-016613 and the Research Group in Physics of the
Hungarian Academy of Sciences, Debrecen.

\def\np#1#2#3  {Nucl.\ Phys.\ {\bf #1}, #2 (19#3)}
\def\pl#1#2#3  {Phys.\ Lett.\ {\bf #1}, #2 (19#3)}
\def\prep#1#2#3  {Phys.\ Rep.\ {\bf #1}, #2 (19#3)}
\def\prd#1#2#3 {Phys.\ Rev.\ D {\bf #1}, #2 (19#3)}
\def\prl#1#2#3 {Phys.\ Rev.\ Lett.\ {\bf #1}, #2 (19#3)}
\def\zpc#1#2#3  {Zeit.\ Phys.\ C {\bf #1}, #2 (19#3)}
\def\cmc#1#2#3  {Comp.\ Phys.\ Comm.\ {\bf #1}, #2 (19#3)}

\section{Appendix}

%\onecolumn
\begin{table}             
\caption{Next-to-leading order kinematical functions to the
$\cos\KSW$ angular distribution.  The Durham jet algorithm is used.}
\begin{tabular}{crrrrrrr}
 $\cos\KSW$& $C_4\quad$ & $C_0\quad$ & $C_x\quad$ & $C_y\quad$ & $C_{xx}\quad$ & $C_{xy}\quad$ & $C_{yy}\quad$     \\   
\tableline                        
-0.950 & \res{ 1190.6}{ 25.0} & \res{ -159.3}{  6.7} & \res{  655.2}{  8.9} & \res{-1504.8}{ 11.8} & \res{  129.7}{  3.6} & \res{ -235.6}{  5.2} & \res{ -128.0}{  0.7} \\
-0.850 & \res{  686.5}{ 21.2} & \res{  -86.7}{  5.8} & \res{  381.8}{  7.6} & \res{ -889.8}{  9.3} & \res{   70.7}{  2.9} & \res{ -109.6}{  3.8} & \res{ -127.9}{  1.0} \\
-0.750 & \res{  696.9}{ 19.4} & \res{  -67.5}{  5.1} & \res{  353.1}{  7.2} & \res{ -826.6}{  9.8} & \res{   73.3}{  2.8} & \res{  -82.4}{  3.6} & \res{ -157.4}{  1.4} \\
-0.650 & \res{  707.5}{ 22.6} & \res{  -73.8}{  5.3} & \res{  352.2}{  7.8} & \res{ -822.8}{  9.7} & \res{   73.9}{  3.0} & \res{  -61.5}{  3.8} & \res{ -179.4}{  1.7} \\
-0.550 & \res{  722.0}{ 20.9} & \res{  -73.7}{  5.0} & \res{  366.5}{  8.1} & \res{ -851.9}{  9.9} & \res{   70.9}{  2.7} & \res{  -48.2}{  3.9} & \res{ -196.7}{  1.8} \\
-0.450 & \res{  716.2}{ 19.7} & \res{  -77.9}{  5.2} & \res{  373.9}{  8.2} & \res{ -881.7}{ 11.0} & \res{   67.0}{  2.4} & \res{  -32.6}{  3.9} & \res{ -203.8}{  2.1} \\
-0.350 & \res{  708.5}{ 19.6} & \res{  -87.8}{  5.9} & \res{  390.4}{  7.9} & \res{ -893.5}{ 10.4} & \res{   57.8}{  2.3} & \res{  -10.3}{  3.5} & \res{ -206.5}{  2.0} \\
-0.250 & \res{  763.1}{ 20.6} & \res{  -78.2}{  6.5} & \res{  418.1}{  9.2} & \res{ -959.0}{ 11.7} & \res{   57.5}{  2.4} & \res{   -2.0}{  3.3} & \res{ -207.1}{  1.9} \\
-0.150 & \res{  752.5}{ 19.7} & \res{  -77.7}{ 14.0} & \res{  435.4}{ 10.6} & \res{ -982.8}{ 11.7} & \res{   46.9}{  2.2} & \res{   10.4}{  3.0} & \res{ -197.2}{  1.8} \\
-0.050 & \res{  730.7}{ 18.8} & \res{ -104.0}{  7.3} & \res{  457.3}{  9.7} & \res{-1039.0}{ 12.7} & \res{   41.1}{  2.1} & \res{   16.1}{  2.6} & \res{ -191.6}{  2.0} \\
 0.050 & \res{  665.5}{ 18.7} & \res{  -90.7}{  7.3} & \res{  420.7}{  9.9} & \res{ -934.2}{ 12.6} & \res{   31.6}{  2.1} & \res{   30.0}{  2.5} & \res{ -176.1}{  1.8} \\
 0.150 & \res{  652.9}{ 18.0} & \res{  -84.7}{  7.2} & \res{  431.0}{  9.6} & \res{ -935.4}{ 13.3} & \res{   21.2}{  2.1} & \res{   40.6}{  2.6} & \res{ -157.4}{  1.5} \\
 0.250 & \res{  631.6}{ 17.0} & \res{  -86.2}{  7.5} & \res{  424.2}{  9.8} & \res{ -904.9}{ 13.4} & \res{   17.5}{  2.1} & \res{   40.2}{  2.1} & \res{ -147.9}{  1.4} \\
 0.350 & \res{  633.7}{ 15.6} & \res{  -83.3}{  6.7} & \res{  433.1}{  9.5} & \res{ -929.5}{ 13.4} & \res{   14.5}{  2.2} & \res{   44.2}{  2.2} & \res{ -137.6}{  1.3} \\
 0.450 & \res{  576.9}{ 29.9} & \res{ -105.1}{  7.3} & \res{  404.6}{ 42.6} & \res{ -923.4}{ 12.7} & \res{   18.9}{ 13.3} & \res{   47.5}{  2.2} & \res{ -127.2}{  1.2} \\
 0.550 & \res{  628.8}{ 29.5} & \res{ -100.3}{  8.5} & \res{  484.4}{ 42.6} & \res{ -980.6}{ 13.5} & \res{   -4.0}{ 13.5} & \res{   52.6}{  2.6} & \res{ -122.8}{  1.1} \\
 0.650 & \res{  628.4}{ 16.2} & \res{ -114.0}{  8.7} & \res{  483.6}{ 12.3} & \res{-1041.5}{ 14.2} & \res{    2.3}{  3.3} & \res{   59.3}{  2.9} & \res{ -119.7}{  1.0} \\
 0.750 & \res{  711.1}{ 15.9} & \res{ -132.8}{  9.8} & \res{  578.1}{ 13.9} & \res{-1197.2}{ 15.5} & \res{   -9.8}{  3.7} & \res{   68.2}{  3.1} & \res{ -118.2}{  0.9} \\
 0.850 & \res{  836.8}{ 15.6} & \res{ -169.5}{ 10.8} & \res{  675.8}{ 14.2} & \res{-1444.1}{ 17.4} & \res{   -3.3}{  4.1} & \res{   74.0}{  3.7} & \res{ -130.3}{  1.1} \\
 0.950 & \res{ 1820.0}{ 22.9} & \res{ -399.8}{ 12.9} & \res{ 1547.2}{ 16.8} & \res{-3258.7}{ 23.5} & \res{  -29.1}{  4.8} & \res{  168.9}{  5.1} & \res{ -252.6}{  1.3} \\
\end{tabular}                                               
\end{table}
\noindent

\begin{table}             
\caption{Next-to-leading order kinematical functions to the
$|\cos\NR|$ angular distribution.  The Durham jet algorithm is used.}
\begin{tabular}{crrrrrrr}
 $|\cos\NR|$& $C_4\quad$ & $C_0\quad$ & $C_x\quad$ & $C_y\quad$ & $C_{xx}\quad$ & $C_{xy}\quad$ & $C_{yy}\quad$     \\   
\tableline                        
 0.025 & \res{ 1184.8}{ 26.6} & \res{ -164.6}{ 10.5} & \res{  757.6}{ 16.8} & \res{-1769.9}{ 24.6} & \res{   46.2}{  5.5} & \res{  161.7}{  7.8} & \res{ -459.4}{  4.3} \\
 0.075 & \res{ 1130.0}{ 35.3} & \res{ -170.6}{ 12.3} & \res{  757.8}{ 21.3} & \res{-1755.2}{ 26.9} & \res{   36.7}{  6.7} & \res{  156.8}{  8.6} & \res{ -450.0}{  4.4} \\
 0.125 & \res{ 1204.4}{ 37.0} & \res{ -164.8}{ 12.0} & \res{  780.8}{ 22.2} & \res{-1783.0}{ 25.5} & \res{   43.0}{  7.1} & \res{  150.8}{  8.2} & \res{ -451.7}{  4.1} \\
 0.175 & \res{ 1190.0}{ 39.9} & \res{ -170.9}{ 14.7} & \res{  764.6}{ 20.2} & \res{-1813.6}{ 28.1} & \res{   52.0}{  6.6} & \res{  142.6}{  8.5} & \res{ -451.4}{  4.1} \\
 0.225 & \res{ 1239.0}{ 37.9} & \res{ -172.1}{ 13.9} & \res{  801.1}{ 18.7} & \res{-1828.6}{ 28.7} & \res{   47.1}{  5.9} & \res{  139.5}{  7.5} & \res{ -432.3}{  4.0} \\
 0.275 & \res{ 1247.5}{ 32.2} & \res{ -172.4}{ 12.4} & \res{  796.1}{ 18.3} & \res{-1832.1}{ 25.7} & \res{   54.2}{  5.8} & \res{  115.2}{  8.0} & \res{ -423.6}{  4.0} \\
 0.325 & \res{ 1268.2}{ 33.5} & \res{ -189.6}{ 15.3} & \res{  828.9}{ 19.7} & \res{-1885.8}{ 25.7} & \res{   54.7}{  6.4} & \res{  100.1}{  8.5} & \res{ -408.9}{  3.5} \\
 0.375 & \res{ 1311.9}{ 40.1} & \res{ -187.4}{ 14.6} & \res{  824.8}{ 20.1} & \res{-1903.0}{ 26.0} & \res{   68.5}{  6.3} & \res{   74.3}{  7.7} & \res{ -386.8}{  3.2} \\
 0.425 & \res{ 1358.3}{ 41.6} & \res{ -177.3}{ 13.0} & \res{  888.3}{ 20.6} & \res{-1986.6}{ 26.8} & \res{   56.5}{  6.3} & \res{   55.9}{  7.4} & \res{ -374.3}{  3.2} \\
 0.475 & \res{ 1456.3}{ 38.0} & \res{ -201.0}{ 13.5} & \res{  916.8}{ 20.5} & \res{-2044.7}{ 27.4} & \res{   74.9}{  5.9} & \res{   38.5}{  7.4} & \res{ -356.7}{  3.0} \\
 0.525 & \res{ 1443.4}{ 41.3} & \res{ -225.8}{ 14.3} & \res{  951.0}{ 20.6} & \res{-2130.4}{ 26.3} & \res{   69.0}{  5.8} & \res{   34.1}{  6.9} & \res{ -342.0}{  3.1} \\
 0.575 & \res{ 1570.0}{ 60.5} & \res{ -208.6}{ 14.3} & \res{  980.2}{ 29.9} & \res{-2198.9}{ 26.5} & \res{   89.4}{  6.3} & \res{  -11.4}{  7.2} & \res{ -317.0}{  2.5} \\
 0.625 & \res{ 1662.7}{ 59.7} & \res{ -249.2}{ 14.6} & \res{ 1085.3}{ 28.5} & \res{-2305.5}{ 26.3} & \res{   79.1}{  5.8} & \res{  -30.2}{  6.7} & \res{ -300.8}{  2.5} \\
 0.675 & \res{ 1806.9}{ 60.9} & \res{ -233.2}{ 15.9} & \res{ 1146.4}{ 30.3} & \res{-2448.1}{ 31.4} & \res{   92.6}{  6.1} & \res{  -61.2}{  7.1} & \res{ -276.8}{  2.2} \\
 0.725 & \res{ 1689.3}{ 62.1} & \res{ -279.2}{ 18.5} & \res{ 1129.5}{ 30.7} & \res{-2503.3}{ 30.8} & \res{   93.3}{  6.3} & \res{  -83.3}{  6.8} & \res{ -249.5}{  1.8} \\
 0.775 & \res{ 1913.3}{ 50.4} & \res{ -263.6}{ 16.6} & \res{ 1244.8}{ 23.8} & \res{-2605.8}{ 29.0} & \res{   93.6}{  6.9} & \res{ -105.4}{  6.2} & \res{ -226.0}{  1.6} \\
 0.825 & \res{ 1931.5}{ 52.0} & \res{ -290.4}{ 17.1} & \res{ 1269.4}{ 22.1} & \res{-2670.3}{ 28.0} & \res{  101.2}{  6.2} & \res{ -139.6}{  6.7} & \res{ -200.1}{  1.4} \\
 0.875 & \res{ 1907.6}{ 50.7} & \res{ -284.7}{ 17.9} & \res{ 1254.5}{ 22.7} & \res{-2691.0}{ 30.5} & \res{  106.3}{  5.7} & \res{ -160.8}{  7.6} & \res{ -173.4}{  1.3} \\
 0.925 & \res{ 1979.0}{ 56.1} & \res{ -302.0}{ 31.4} & \res{ 1297.6}{ 26.1} & \res{-2777.3}{ 29.1} & \res{  116.5}{  6.3} & \res{ -196.4}{  7.4} & \res{ -151.9}{  1.1} \\
 0.975 & \res{ 2426.4}{ 68.3} & \res{ -399.1}{ 19.4} & \res{ 1658.0}{ 27.8} & \res{-3469.3}{ 33.7} & \res{  122.2}{  6.5} & \res{ -241.3}{  8.3} & \res{ -137.8}{  0.8} \\
\end{tabular}                                               
\end{table}
\noindent

\begin{table}             
\caption{Next-to-leading order kinematical functions to the
$\cos\jetalpha$ angular distribution. The Durham jet algorithm is used.}
\begin{tabular}{crrrrrrr}
 $\cos\jetalpha$& $C_4\quad$ & $C_0\quad$ & $C_x\quad$ & $C_y\quad$ & $C_{xx}\quad$ & $C_{xy}\quad$ & $C_{yy}\quad$     \\   
\tableline                        
-0.950 & \res{ 1038.4}{ 17.9} & \res{ -227.3}{  9.1} & \res{  901.2}{ 12.1} & \res{-1888.6}{ 17.5} & \res{  -20.6}{  2.8} & \res{   77.1}{  2.9} & \res{ -106.0}{  0.8} \\
-0.850 & \res{  988.9}{ 19.5} & \res{ -187.7}{  9.0} & \res{  830.9}{ 12.8} & \res{-1717.5}{ 17.9} & \res{  -19.0}{  3.0} & \res{   73.8}{  2.9} & \res{ -109.0}{  0.8} \\
-0.750 & \res{  948.7}{ 19.4} & \res{ -177.0}{  9.2} & \res{  777.1}{ 13.9} & \res{-1628.8}{ 16.6} & \res{  -10.4}{  3.5} & \res{   67.2}{  2.7} & \res{ -116.9}{  0.9} \\
-0.650 & \res{  911.1}{ 22.2} & \res{ -162.4}{  9.6} & \res{  716.5}{ 13.0} & \res{-1500.5}{ 15.4} & \res{   -0.9}{  2.9} & \res{   54.8}{  2.7} & \res{ -123.8}{  1.0} \\
-0.550 & \res{  899.6}{ 26.1} & \res{ -144.1}{  7.9} & \res{  684.6}{ 14.7} & \res{-1427.9}{ 15.6} & \res{    3.4}{  3.1} & \res{   47.9}{  2.7} & \res{ -132.5}{  1.0} \\
-0.450 & \res{  849.9}{ 22.1} & \res{ -131.0}{ 15.0} & \res{  629.7}{ 14.3} & \res{-1381.1}{ 15.8} & \res{   12.9}{  3.4} & \res{   43.0}{  3.0} & \res{ -141.0}{  1.1} \\
-0.350 & \res{  827.5}{ 21.1} & \res{ -133.5}{  7.3} & \res{  597.8}{ 11.5} & \res{-1268.9}{ 15.7} & \res{   16.0}{  3.2} & \res{   38.1}{  2.9} & \res{ -148.8}{  1.2} \\
-0.250 & \res{  816.0}{ 19.0} & \res{ -126.8}{  6.8} & \res{  566.3}{ 11.6} & \res{-1235.8}{ 14.7} & \res{   25.9}{  2.9} & \res{   26.5}{  2.8} & \res{ -158.2}{  1.2} \\
-0.150 & \res{  810.4}{ 19.8} & \res{ -125.3}{  7.0} & \res{  546.5}{ 12.2} & \res{-1200.7}{ 14.2} & \res{   32.2}{  3.6} & \res{   20.0}{  3.2} & \res{ -171.5}{  1.3} \\
-0.050 & \res{  843.6}{ 21.4} & \res{ -118.9}{  7.4} & \res{  534.6}{ 13.5} & \res{-1187.5}{ 14.9} & \res{   44.4}{  3.9} & \res{    7.4}{  3.2} & \res{ -187.6}{  1.5} \\
 0.050 & \res{  829.9}{ 21.7} & \res{ -126.0}{  6.7} & \res{  510.9}{ 10.3} & \res{-1152.5}{ 13.3} & \res{   53.0}{  2.7} & \res{   -1.4}{  3.4} & \res{ -203.9}{  1.6} \\
 0.150 & \res{  879.4}{ 23.4} & \res{ -105.2}{  7.2} & \res{  484.1}{ 10.1} & \res{-1100.1}{ 13.4} & \res{   69.4}{  2.9} & \res{  -14.5}{  3.9} & \res{ -219.1}{  1.8} \\
 0.250 & \res{  834.2}{ 21.6} & \res{ -104.4}{  6.2} & \res{  450.9}{  8.9} & \res{-1055.2}{ 12.4} & \res{   74.0}{  2.7} & \res{  -28.7}{  4.2} & \res{ -233.1}{  2.1} \\
 0.350 & \res{  837.2}{ 23.9} & \res{  -93.4}{  5.6} & \res{  426.1}{  8.7} & \res{ -999.6}{ 11.9} & \res{   83.0}{  3.1} & \res{  -43.0}{  4.5} & \res{ -244.0}{  2.0} \\
 0.450 & \res{  828.0}{ 24.0} & \res{  -79.6}{  5.3} & \res{  382.7}{  7.8} & \res{ -915.3}{ 11.1} & \res{   93.2}{  3.4} & \res{  -55.2}{  4.9} & \res{ -249.9}{  2.1} \\
 0.550 & \res{  793.7}{ 22.2} & \res{  -72.9}{  4.5} & \res{  343.9}{  8.9} & \res{ -854.5}{ 10.3} & \res{  100.1}{  2.7} & \res{  -70.2}{  5.0} & \res{ -252.9}{  2.1} \\
 0.650 & \res{  719.8}{ 21.9} & \res{  -59.1}{  5.0} & \res{  305.2}{  8.4} & \res{ -763.9}{  8.4} & \res{   94.0}{  2.9} & \res{  -74.7}{  5.0} & \res{ -238.4}{  2.1} \\
 0.750 & \res{  546.0}{ 15.0} & \res{  -52.5}{  4.1} & \res{  238.4}{  5.2} & \res{ -602.4}{  7.5} & \res{   73.0}{  2.5} & \res{  -66.1}{  4.0} & \res{ -184.8}{  2.1} \\
 0.850 & \res{  245.2}{  9.3} & \res{  -24.0}{  2.9} & \res{  130.1}{  3.4} & \res{ -302.8}{  5.2} & \res{   24.8}{  1.7} & \res{  -31.7}{  2.4} & \res{  -62.8}{  0.9} \\
 0.950 & \res{   12.5}{  2.4} & \res{   -1.9}{  1.0} & \res{    9.2}{  1.3} & \res{  -17.7}{  1.4} & \res{    0.2}{  0.2} & \res{   -0.4}{  0.2} & \res{   -0.9}{  0.1} \\
\end{tabular}                                               
\end{table}
\noindent

\begin{table}             
\caption{Next-to-leading order kinematical functions to the
$|\cos\BZ|$ angular distribution. The Durham jet algorithm is used.}
\begin{tabular}{crrrrrrr}
 $|\cos\BZ|$& $C_4\quad$ & $C_0\quad$ & $C_x\quad$ & $C_y\quad$ & $C_{xx}\quad$ & $C_{xy}\quad$ & $C_{yy}\quad$     \\   
\tableline                        
 0.025 & \res{ 1133.9}{ 39.7} & \res{ -171.8}{ 14.1} & \res{  748.6}{ 24.7} & \res{-1623.5}{ 30.9} & \res{   29.2}{  6.3} & \res{  161.8}{  8.2} & \res{ -400.9}{  4.4} \\
 0.075 & \res{ 1087.1}{ 47.9} & \res{ -170.2}{ 17.3} & \res{  724.3}{ 35.3} & \res{-1689.3}{ 31.6} & \res{   34.9}{  9.9} & \res{  164.9}{  9.2} & \res{ -395.9}{  4.2} \\
 0.125 & \res{ 1147.7}{ 49.5} & \res{ -163.0}{ 18.2} & \res{  771.0}{ 35.7} & \res{-1640.7}{ 26.6} & \res{   23.1}{ 11.0} & \res{  153.2}{  7.8} & \res{ -395.1}{  4.1} \\
 0.175 & \res{ 1075.4}{ 45.9} & \res{ -160.2}{ 16.6} & \res{  719.1}{ 27.9} & \res{-1665.6}{ 26.8} & \res{   33.6}{  8.9} & \res{  159.0}{  9.3} & \res{ -391.4}{  4.2} \\
 0.225 & \res{ 1144.5}{ 42.5} & \res{ -177.8}{ 16.9} & \res{  752.8}{ 27.3} & \res{-1659.9}{ 25.7} & \res{   36.9}{  7.5} & \res{  135.4}{  8.9} & \res{ -381.4}{  3.9} \\
 0.275 & \res{ 1110.3}{ 43.5} & \res{ -172.6}{ 17.9} & \res{  741.9}{ 29.0} & \res{-1634.2}{ 27.2} & \res{   33.4}{  8.2} & \res{  121.7}{  8.1} & \res{ -361.1}{  3.7} \\
 0.325 & \res{ 1152.2}{ 42.3} & \res{ -142.2}{ 18.6} & \res{  744.5}{ 28.2} & \res{-1651.3}{ 26.3} & \res{   39.0}{  7.8} & \res{  110.2}{  7.7} & \res{ -350.3}{  3.1} \\
 0.375 & \res{ 1181.2}{ 43.1} & \res{ -175.4}{ 16.9} & \res{  794.6}{ 28.2} & \res{-1741.2}{ 26.3} & \res{   35.9}{  7.3} & \res{  106.1}{  6.8} & \res{ -340.5}{  3.3} \\
 0.425 & \res{ 1210.8}{ 47.8} & \res{ -155.7}{ 16.7} & \res{  798.5}{ 36.3} & \res{-1693.3}{ 26.2} & \res{   37.9}{ 10.1} & \res{   72.5}{  7.5} & \res{ -330.6}{  3.2} \\
 0.475 & \res{ 1213.1}{ 44.5} & \res{ -185.8}{ 17.7} & \res{  762.7}{ 34.5} & \res{-1749.9}{ 25.3} & \res{   66.2}{  9.4} & \res{   56.7}{  7.7} & \res{ -322.3}{  2.9} \\
 0.525 & \res{ 1236.6}{ 38.9} & \res{ -195.1}{ 18.2} & \res{  819.6}{ 24.7} & \res{-1801.7}{ 23.2} & \res{   53.0}{  6.6} & \res{   44.5}{  6.3} & \res{ -310.7}{  2.8} \\
 0.575 & \res{ 1337.2}{ 42.9} & \res{ -182.4}{ 18.0} & \res{  847.8}{ 24.9} & \res{-1855.7}{ 24.3} & \res{   65.6}{  6.7} & \res{   19.1}{  5.8} & \res{ -299.8}{  2.5} \\
 0.625 & \res{ 1383.5}{ 44.4} & \res{ -201.5}{ 16.0} & \res{  891.7}{ 23.7} & \res{-1983.3}{ 24.3} & \res{   72.4}{  6.5} & \res{   -2.5}{  5.8} & \res{ -292.6}{  2.3} \\
 0.675 & \res{ 1415.8}{ 44.2} & \res{ -197.9}{ 15.9} & \res{  916.5}{ 24.2} & \res{-2050.8}{ 25.1} & \res{   74.8}{  6.4} & \res{  -22.5}{  6.1} & \res{ -279.9}{  2.0} \\
 0.725 & \res{ 1551.9}{ 44.0} & \res{ -283.2}{ 89.5} & \res{ 1039.4}{ 77.4} & \res{-2150.5}{ 25.3} & \res{   74.9}{ 15.7} & \res{  -45.4}{  6.4} & \res{ -268.6}{  2.0} \\
 0.775 & \res{ 1659.4}{ 47.8} & \res{ -152.9}{ 90.5} & \res{ 1020.2}{ 77.7} & \res{-2357.4}{ 26.3} & \res{   99.2}{ 15.8} & \res{  -75.9}{  6.7} & \res{ -262.1}{  1.9} \\
 0.825 & \res{ 1828.2}{ 48.2} & \res{ -251.4}{ 17.7} & \res{ 1166.8}{ 26.1} & \res{-2553.8}{ 25.9} & \res{  105.5}{  6.3} & \res{ -102.4}{  7.0} & \res{ -248.7}{  1.6} \\
 0.875 & \res{ 1791.5}{245.1} & \res{ -316.9}{ 32.0} & \res{ 1236.4}{ 91.5} & \res{-2863.7}{ 27.2} & \res{  110.1}{  8.8} & \res{ -146.7}{  7.0} & \res{ -243.1}{  1.6} \\
 0.925 & \res{ 2455.5}{246.3} & \res{ -318.9}{ 33.3} & \res{ 1528.1}{ 91.8} & \res{-3230.3}{ 28.2} & \res{  151.2}{  9.1} & \res{ -217.0}{  7.9} & \res{ -249.3}{  1.5} \\
 0.975 & \res{ 4804.5}{ 70.2} & \res{ -731.4}{ 36.2} & \res{ 3109.1}{ 34.0} & \res{-6806.4}{ 41.7} & \res{  320.2}{  8.8} & \res{ -552.9}{ 12.4} & \res{ -445.9}{  1.9} \\
\end{tabular}                                               
\end{table}
\noindent

\begin{table}             
\caption{Next-to-leading order kinematical functions to the
$\cos\KSW$ angular distribution. The Cambridge jet algorithm is used.}
\begin{tabular}{crrrrrrr}
 $\cos\KSW$& $C_4\quad$ & $C_0\quad$ & $C_x\quad$ & $C_y\quad$ & $C_{xx}\quad$ & $C_{xy}\quad$ & $C_{yy}\quad$     \\  
\tableline                        
-0.950 & \res{ 1085.9}{ 44.0} & \res{ -287.0}{  8.1} & \res{  713.0}{ 12.8} & \res{-1754.9}{ 17.4} & \res{  135.1}{  7.2} & \res{ -280.6}{  8.0} & \res{ -143.7}{  0.9} \\
-0.850 & \res{  658.4}{ 35.4} & \res{ -164.3}{  7.1} & \res{  404.4}{ 10.0} & \res{ -986.2}{ 12.6} & \res{   81.7}{  4.9} & \res{ -134.9}{  6.1} & \res{ -134.0}{  1.2} \\
-0.750 & \res{  617.4}{ 27.0} & \res{ -139.8}{  8.8} & \res{  366.6}{  9.6} & \res{ -932.2}{ 13.9} & \res{   78.5}{  3.7} & \res{ -108.1}{  6.0} & \res{ -170.1}{  1.6} \\
-0.650 & \res{  620.0}{ 23.2} & \res{ -124.2}{  8.8} & \res{  355.9}{  9.5} & \res{ -924.0}{ 14.5} & \res{   74.7}{  3.1} & \res{  -71.1}{  5.7} & \res{ -196.1}{  2.0} \\
-0.550 & \res{  662.8}{ 23.0} & \res{ -130.5}{  7.9} & \res{  377.7}{ 10.5} & \res{ -961.9}{ 13.0} & \res{   75.7}{  3.2} & \res{  -59.5}{  4.6} & \res{ -210.4}{  2.2} \\
-0.450 & \res{  605.2}{ 24.4} & \res{ -145.8}{  8.2} & \res{  380.8}{ 10.6} & \res{ -954.7}{ 13.1} & \res{   60.8}{  3.0} & \res{  -30.6}{  4.4} & \res{ -216.4}{  2.5} \\
-0.350 & \res{  658.4}{ 23.8} & \res{ -143.1}{  7.8} & \res{  405.4}{ 10.4} & \res{-1004.1}{ 14.8} & \res{   60.9}{  2.7} & \res{  -13.3}{  4.4} & \res{ -224.1}{  2.7} \\
-0.250 & \res{  653.0}{ 25.4} & \res{ -148.1}{  7.6} & \res{  421.2}{ 11.6} & \res{-1035.8}{ 15.6} & \res{   54.3}{  2.6} & \res{   -4.2}{  4.6} & \res{ -219.6}{  2.5} \\
-0.150 & \res{  610.7}{ 25.0} & \res{ -173.7}{  8.8} & \res{  431.9}{ 12.1} & \res{-1044.3}{ 16.1} & \res{   44.3}{  2.8} & \res{   13.4}{  4.3} & \res{ -213.7}{  2.4} \\
-0.050 & \res{  603.3}{ 23.6} & \res{ -176.6}{ 12.3} & \res{  449.2}{ 15.2} & \res{-1078.7}{ 16.5} & \res{   36.9}{  3.4} & \res{   16.1}{  3.8} & \res{ -192.3}{  2.2} \\
 0.050 & \res{  548.4}{ 21.4} & \res{ -147.6}{  7.5} & \res{  401.7}{ 10.6} & \res{ -970.0}{ 14.6} & \res{   30.8}{  2.5} & \res{   28.0}{  2.6} & \res{ -171.0}{  2.0} \\
 0.150 & \res{  514.7}{ 19.3} & \res{ -161.5}{  7.3} & \res{  414.3}{ 10.2} & \res{ -952.0}{ 13.2} & \res{   18.2}{  2.4} & \res{   37.2}{  2.9} & \res{ -161.0}{  1.8} \\
 0.250 & \res{  508.4}{ 17.1} & \res{ -153.8}{  6.9} & \res{  407.1}{ 10.0} & \res{ -942.3}{ 12.7} & \res{   17.3}{  2.2} & \res{   38.5}{  2.9} & \res{ -147.1}{  1.6} \\
 0.350 & \res{  500.3}{ 17.0} & \res{ -158.0}{  7.4} & \res{  425.3}{ 13.3} & \res{ -974.9}{ 14.6} & \res{    9.9}{  3.2} & \res{   42.6}{  2.2} & \res{ -140.6}{  1.5} \\
 0.450 & \res{  497.7}{ 18.5} & \res{ -164.4}{  7.5} & \res{  429.0}{ 13.9} & \res{ -996.6}{ 13.9} & \res{   10.0}{  3.4} & \res{   44.7}{  2.4} & \res{ -129.7}{  1.3} \\
 0.550 & \res{  471.1}{ 16.9} & \res{ -177.1}{  7.9} & \res{  444.7}{ 12.0} & \res{-1027.1}{ 15.3} & \res{    1.1}{  2.9} & \res{   52.3}{  2.7} & \res{ -124.8}{  1.3} \\
 0.650 & \res{  510.0}{ 16.8} & \res{ -195.1}{  7.8} & \res{  489.6}{ 11.6} & \res{-1085.2}{ 15.0} & \res{   -4.2}{  3.1} & \res{   56.5}{  2.9} & \res{ -120.9}{  1.2} \\
 0.750 & \res{  560.0}{ 17.4} & \res{ -207.1}{  8.3} & \res{  545.7}{ 12.3} & \res{-1217.8}{ 15.2} & \res{   -8.4}{  3.2} & \res{   65.2}{  3.4} & \res{ -118.7}{  1.0} \\
 0.850 & \res{  681.4}{ 22.0} & \res{ -264.5}{  9.5} & \res{  687.5}{ 14.6} & \res{-1514.9}{ 18.8} & \res{  -16.0}{  3.7} & \res{   78.5}{  3.9} & \res{ -129.7}{  1.2} \\
 0.950 & \res{ 1493.1}{ 25.4} & \res{ -630.2}{ 14.1} & \res{ 1574.3}{ 18.8} & \res{-3435.0}{ 24.6} & \res{  -48.4}{  5.7} & \res{  176.7}{  6.1} & \res{ -252.8}{  1.4} \\
\end{tabular}                                               
\end{table}
\noindent

\begin{table}             
\caption{Next-to-leading order kinematical functions to the
$|\cos\NR|$ angular distribution. The Cambridge jet algorithm is used.}
\begin{tabular}{crrrrrrr}
 $|\cos\NR|$& $C_4\quad$ & $C_0\quad$ & $C_x\quad$ & $C_y\quad$ & $C_{xx}\quad$ & $C_{xy}\quad$ & $C_{yy}\quad$     \\   
\tableline                        
 0.025 & \res{ 1016.1}{ 30.2} & \res{ -233.6}{ 14.4} & \res{  750.2}{ 20.8} & \res{-1827.3}{ 28.5} & \res{   34.8}{  6.8} & \res{  160.4}{  9.5} & \res{ -449.3}{  4.6} \\
 0.075 & \res{ 1023.7}{ 38.2} & \res{ -237.2}{ 12.6} & \res{  734.5}{ 22.5} & \res{-1817.0}{ 32.1} & \res{   46.1}{  7.2} & \res{  136.2}{ 10.2} & \res{ -450.5}{  4.6} \\
 0.125 & \res{ 1011.2}{ 44.1} & \res{ -224.1}{ 13.7} & \res{  756.4}{ 23.2} & \res{-1848.0}{ 30.1} & \res{   31.7}{  7.3} & \res{  157.7}{  9.7} & \res{ -441.6}{  5.0} \\
 0.175 & \res{ 1038.5}{ 44.3} & \res{ -249.2}{ 13.3} & \res{  760.8}{ 21.9} & \res{-1867.3}{ 27.9} & \res{   44.8}{  6.8} & \res{  126.8}{ 10.1} & \res{ -450.3}{  4.5} \\
 0.225 & \res{ 1037.3}{ 42.3} & \res{ -240.7}{ 11.0} & \res{  760.2}{ 24.2} & \res{-1867.6}{ 29.5} & \res{   43.9}{  6.7} & \res{  117.8}{ 11.4} & \res{ -430.8}{  4.0} \\
 0.275 & \res{ 1112.1}{ 42.0} & \res{ -253.5}{ 12.1} & \res{  807.4}{ 23.2} & \res{-1937.9}{ 27.2} & \res{   47.7}{  6.8} & \res{  117.0}{ 11.3} & \res{ -426.1}{  4.6} \\
 0.325 & \res{ 1078.7}{ 37.7} & \res{ -260.3}{ 13.6} & \res{  808.9}{ 21.4} & \res{-1937.7}{ 27.0} & \res{   44.1}{  6.7} & \res{   99.2}{  9.0} & \res{ -416.5}{  4.0} \\
 0.375 & \res{ 1074.5}{ 37.2} & \res{ -277.2}{ 13.6} & \res{  797.5}{ 21.5} & \res{-1997.0}{ 29.2} & \res{   57.6}{  6.9} & \res{   87.5}{ 10.1} & \res{ -403.2}{  3.8} \\
 0.425 & \res{ 1147.9}{ 39.6} & \res{ -314.1}{ 15.2} & \res{  896.8}{ 22.6} & \res{-2103.0}{ 28.6} & \res{   46.6}{  6.4} & \res{   57.1}{ 10.0} & \res{ -385.7}{  3.7} \\
 0.475 & \res{ 1222.8}{ 43.0} & \res{ -294.4}{ 13.3} & \res{  911.7}{ 23.0} & \res{-2166.9}{ 30.0} & \res{   58.9}{  6.6} & \res{   38.1}{  7.9} & \res{ -370.6}{  3.9} \\
 0.525 & \res{ 1238.6}{ 46.5} & \res{ -343.9}{ 15.0} & \res{  944.9}{ 23.2} & \res{-2233.5}{ 27.5} & \res{   62.5}{  6.7} & \res{   29.5}{  8.0} & \res{ -352.4}{  3.7} \\
 0.575 & \res{ 1295.5}{ 45.3} & \res{ -351.4}{ 15.1} & \res{  969.0}{ 23.0} & \res{-2346.7}{ 30.8} & \res{   78.8}{  6.0} & \res{   -4.6}{  7.1} & \res{ -334.3}{  3.5} \\
 0.625 & \res{ 1371.8}{ 45.0} & \res{ -393.9}{ 14.4} & \res{ 1069.7}{ 22.8} & \res{-2488.9}{ 31.3} & \res{   72.3}{  6.0} & \res{  -39.0}{  8.7} & \res{ -315.0}{  3.3} \\
 0.675 & \res{ 1403.5}{ 55.5} & \res{ -452.3}{ 17.6} & \res{ 1134.1}{ 25.8} & \res{-2606.4}{ 34.4} & \res{   75.7}{  6.7} & \res{  -73.7}{  8.6} & \res{ -294.4}{  2.8} \\
 0.725 & \res{ 1504.7}{ 61.1} & \res{ -455.9}{ 25.1} & \res{ 1183.5}{ 33.8} & \res{-2732.3}{ 34.7} & \res{   85.6}{  8.9} & \res{  -86.4}{  8.1} & \res{ -274.2}{  2.8} \\
 0.775 & \res{ 1479.6}{ 59.4} & \res{ -539.4}{ 17.8} & \res{ 1202.1}{ 29.3} & \res{-2802.7}{ 33.0} & \res{   99.7}{  8.7} & \res{ -126.3}{  8.0} & \res{ -250.6}{  2.5} \\
 0.825 & \res{ 1527.5}{ 63.3} & \res{ -566.6}{ 20.2} & \res{ 1268.4}{ 32.6} & \res{-2896.6}{ 34.1} & \res{   94.9}{  8.2} & \res{ -142.1}{  8.7} & \res{ -230.8}{  2.4} \\
 0.875 & \res{ 1585.8}{ 64.0} & \res{ -629.1}{ 21.1} & \res{ 1318.0}{ 27.1} & \res{-2971.0}{ 34.6} & \res{  110.0}{  7.3} & \res{ -196.0}{ 10.0} & \res{ -204.7}{  2.0} \\
 0.925 & \res{ 1706.3}{ 87.7} & \res{ -625.3}{ 20.9} & \res{ 1357.8}{ 29.4} & \res{-3096.3}{ 37.4} & \res{  130.8}{ 10.4} & \res{ -234.3}{ 11.1} & \res{ -180.9}{  1.8} \\
 0.975 & \res{ 2243.9}{ 92.8} & \res{ -842.0}{ 23.9} & \res{ 1818.8}{ 32.5} & \res{-4041.0}{ 44.8} & \res{  160.3}{ 13.9} & \res{ -329.7}{ 15.2} & \res{ -171.8}{  1.5} \\
\end{tabular}                                               
\end{table}
\noindent

\begin{table}             
\caption{Next-to-leading order kinematical functions to the
$\cos\jetalpha$ angular distribution. The Cambridge jet algorithm is
used.}
\begin{tabular}{crrrrrrr}
 $\cos\jetalpha$& $C_4\quad$ & $C_0\quad$ & $C_x\quad$ & $C_y\quad$ & $C_{xx}\quad$ & $C_{xy}\quad$ & $C_{yy}\quad$     \\   
\tableline                        
-0.950 & \res{  827.2}{ 21.4} & \res{ -373.0}{  9.4} & \res{  904.6}{ 13.6} & \res{-1956.5}{ 16.9} & \res{  -30.4}{  3.3} & \res{   79.5}{  3.1} & \res{ -105.8}{  0.9} \\
-0.850 & \res{  786.6}{ 41.6} & \res{ -333.1}{ 10.3} & \res{  819.7}{ 20.6} & \res{-1750.8}{ 19.7} & \res{  -22.2}{  3.0} & \res{   70.2}{  3.3} & \res{ -108.9}{  0.9} \\
-0.750 & \res{  731.5}{ 40.2} & \res{ -310.3}{ 11.6} & \res{  760.7}{ 19.6} & \res{-1686.0}{ 17.9} & \res{  -15.0}{  3.2} & \res{   64.5}{  3.2} & \res{ -118.2}{  1.0} \\
-0.650 & \res{  726.5}{ 21.4} & \res{ -285.2}{ 10.0} & \res{  718.2}{ 13.6} & \res{-1607.3}{ 18.0} & \res{   -6.1}{  3.2} & \res{   56.4}{  3.1} & \res{ -126.8}{  1.1} \\
-0.550 & \res{  691.0}{ 30.1} & \res{ -257.9}{  8.5} & \res{  666.9}{ 15.9} & \res{-1473.9}{ 18.3} & \res{   -4.3}{  4.2} & \res{   48.0}{  3.3} & \res{ -133.0}{  1.2} \\
-0.450 & \res{  704.7}{ 32.9} & \res{ -242.1}{  9.5} & \res{  635.0}{ 16.8} & \res{-1439.9}{ 18.6} & \res{    8.5}{  4.2} & \res{   40.2}{  2.8} & \res{ -143.0}{  1.2} \\
-0.350 & \res{  692.8}{ 23.8} & \res{ -211.5}{  9.5} & \res{  590.3}{ 13.9} & \res{-1344.2}{ 21.1} & \res{   14.2}{  3.4} & \res{   35.6}{  5.2} & \res{ -149.0}{  1.3} \\
-0.250 & \res{  690.1}{ 21.9} & \res{ -203.2}{  7.9} & \res{  563.4}{ 12.8} & \res{-1266.4}{ 20.2} & \res{   20.0}{  3.1} & \res{   24.9}{  5.2} & \res{ -158.9}{  1.4} \\
-0.150 & \res{  670.2}{ 22.9} & \res{ -193.3}{  8.8} & \res{  531.8}{ 12.5} & \res{-1250.5}{ 17.0} & \res{   28.9}{  3.0} & \res{   17.4}{  3.4} & \res{ -174.4}{  1.5} \\
-0.050 & \res{  699.2}{ 23.9} & \res{ -180.2}{  8.3} & \res{  511.6}{ 12.3} & \res{-1212.6}{ 16.9} & \res{   38.8}{  3.1} & \res{   13.9}{  3.7} & \res{ -185.9}{  1.9} \\
 0.050 & \res{  720.4}{ 30.1} & \res{ -176.5}{  7.3} & \res{  499.8}{ 13.8} & \res{-1163.4}{ 14.7} & \res{   48.0}{  2.8} & \res{   -7.5}{  4.1} & \res{ -202.7}{  1.8} \\
 0.150 & \res{  769.0}{ 31.7} & \res{ -158.3}{  7.4} & \res{  487.1}{ 14.7} & \res{-1185.0}{ 16.3} & \res{   62.2}{  3.0} & \res{  -10.5}{  4.7} & \res{ -223.5}{  2.1} \\
 0.250 & \res{  791.6}{ 33.7} & \res{ -163.2}{  7.5} & \res{  477.7}{ 11.2} & \res{-1163.8}{ 15.8} & \res{   74.3}{  4.3} & \res{  -31.5}{  4.9} & \res{ -239.8}{  2.1} \\
 0.350 & \res{  790.2}{ 28.0} & \res{ -159.0}{  6.8} & \res{  451.3}{ 10.4} & \res{-1119.0}{ 14.8} & \res{   86.1}{  3.6} & \res{  -54.3}{  5.6} & \res{ -260.3}{  2.3} \\
 0.450 & \res{  816.2}{ 33.8} & \res{ -143.8}{  6.7} & \res{  426.7}{ 11.2} & \res{-1095.9}{ 14.4} & \res{  100.2}{  4.2} & \res{  -69.3}{  5.8} & \res{ -273.9}{  2.9} \\
 0.550 & \res{  733.4}{ 30.9} & \res{ -148.9}{  7.1} & \res{  370.3}{ 10.9} & \res{-1033.3}{ 14.7} & \res{  110.4}{  4.2} & \res{  -98.8}{  7.0} & \res{ -277.3}{  2.7} \\
 0.650 & \res{  634.0}{ 35.4} & \res{ -143.2}{  6.3} & \res{  333.7}{  9.4} & \res{ -942.1}{ 14.4} & \res{   98.3}{  5.9} & \res{  -97.3}{  7.1} & \res{ -263.6}{  2.7} \\
 0.750 & \res{  434.4}{ 22.9} & \res{ -122.8}{ 10.4} & \res{  235.7}{ 13.3} & \res{ -732.8}{ 11.0} & \res{   80.7}{  4.8} & \res{  -91.5}{  5.9} & \res{ -201.8}{  2.4} \\
 0.850 & \res{  150.1}{ 14.0} & \res{  -79.6}{  3.6} & \res{  135.6}{  7.6} & \res{ -355.6}{  8.4} & \res{   20.2}{  3.0} & \res{  -41.5}{  4.7} & \res{  -69.4}{  1.4} \\
 0.950 & \res{    0.9}{  1.7} & \res{   -7.2}{  1.0} & \res{    5.1}{  1.0} & \res{  -13.8}{  1.3} & \res{    0.5}{  0.2} & \res{   -0.9}{  0.2} & \res{   -0.7}{  0.1} \\
\end{tabular}                                               
\end{table}
\noindent

\begin{table}             
\caption{Next-to-leading order kinematical functions to the
$|\cos\BZ|$ angular distribution. The Cambridge jet algorithm is used.}
\begin{tabular}{crrrrrrr}
 $|\cos\BZ|$& $C_4\quad$ & $C_0\quad$ & $C_x\quad$ & $C_y\quad$ & $C_{xx}\quad$ & $C_{xy}\quad$ & $C_{yy}\quad$     \\   
\tableline                        
 0.025 & \res{  950.6}{ 40.9} & \res{ -233.8}{ 16.6} & \res{  718.7}{ 29.8} & \res{-1751.2}{ 31.7} & \res{   28.4}{  8.0} & \res{  156.9}{  8.1} & \res{ -398.6}{  4.5} \\
 0.075 & \res{  968.5}{ 53.7} & \res{ -242.8}{ 16.5} & \res{  762.8}{ 35.1} & \res{-1695.1}{ 30.3} & \res{   11.9}{  9.3} & \res{  151.7}{  8.6} & \res{ -395.7}{  4.9} \\
 0.125 & \res{  949.3}{ 52.6} & \res{ -245.5}{ 16.7} & \res{  718.9}{ 34.7} & \res{-1784.5}{ 34.0} & \res{   31.7}{  9.0} & \res{  168.0}{  9.6} & \res{ -404.9}{  5.2} \\
 0.175 & \res{  989.3}{ 45.2} & \res{ -246.1}{ 17.8} & \res{  751.7}{ 32.2} & \res{-1713.5}{ 32.8} & \res{   22.8}{  9.2} & \res{  141.4}{  8.4} & \res{ -390.4}{  4.6} \\
 0.225 & \res{  959.7}{ 45.3} & \res{ -260.6}{ 21.7} & \res{  722.5}{ 35.1} & \res{-1707.9}{ 32.1} & \res{   34.8}{  9.7} & \res{  138.6}{  8.3} & \res{ -383.3}{  4.7} \\
 0.275 & \res{  940.8}{ 45.5} & \res{ -288.5}{ 38.1} & \res{  736.2}{ 38.3} & \res{-1702.2}{ 29.2} & \res{   33.1}{  8.6} & \res{  114.1}{  8.2} & \res{ -373.5}{  4.4} \\
 0.325 & \res{  989.1}{ 48.3} & \res{ -224.8}{ 39.8} & \res{  710.5}{ 38.1} & \res{-1744.9}{ 29.7} & \res{   45.0}{  9.1} & \res{  108.5}{  8.7} & \res{ -367.0}{  4.3} \\
 0.375 & \res{ 1011.7}{ 52.2} & \res{ -257.5}{ 17.5} & \res{  781.7}{ 30.0} & \res{-1804.6}{ 31.9} & \res{   30.5}{  8.2} & \res{   96.4}{  9.3} & \res{ -358.6}{  4.0} \\
 0.425 & \res{ 1055.0}{ 49.7} & \res{ -262.7}{ 17.0} & \res{  767.8}{ 30.7} & \res{-1870.0}{ 31.5} & \res{   55.2}{  8.3} & \res{   75.1}{  9.0} & \res{ -345.1}{  3.3} \\
 0.475 & \res{ 1015.4}{ 71.3} & \res{ -378.7}{ 99.1} & \res{  862.4}{ 63.3} & \res{-1902.1}{ 32.4} & \res{   31.6}{ 15.2} & \res{   63.9}{  8.4} & \res{ -344.3}{  3.8} \\
 0.525 & \res{ 1110.7}{ 89.5} & \res{ -194.5}{ 98.2} & \res{  764.8}{ 68.1} & \res{-1943.5}{ 32.6} & \res{   63.6}{ 15.0} & \res{   46.9}{  7.7} & \res{ -328.1}{  3.4} \\
 0.575 & \res{ 1136.4}{ 75.5} & \res{ -297.7}{ 18.6} & \res{  834.4}{ 39.9} & \res{-1991.1}{ 29.0} & \res{   67.3}{  8.0} & \res{    8.2}{  9.6} & \res{ -308.5}{  3.0} \\
 0.625 & \res{ 1156.8}{ 50.0} & \res{ -313.0}{ 18.6} & \res{  893.3}{ 27.0} & \res{-2106.2}{ 28.9} & \res{   58.5}{  7.7} & \res{   -7.5}{  9.7} & \res{ -311.5}{  3.5} \\
 0.675 & \res{ 1223.0}{ 60.9} & \res{ -379.1}{ 17.0} & \res{  947.1}{ 33.5} & \res{-2229.2}{ 28.1} & \res{   74.0}{  7.6} & \res{  -31.0}{  7.4} & \res{ -298.0}{  2.7} \\
 0.725 & \res{ 1257.6}{ 65.3} & \res{ -391.6}{ 27.5} & \res{ 1005.7}{ 40.8} & \res{-2386.4}{ 30.4} & \res{   72.3}{ 10.5} & \res{  -51.3}{  8.4} & \res{ -288.4}{  2.6} \\
 0.775 & \res{ 1355.8}{ 56.8} & \res{ -435.0}{ 27.6} & \res{ 1058.5}{ 34.6} & \res{-2530.1}{ 28.3} & \res{   93.4}{  9.7} & \res{  -92.2}{  8.9} & \res{ -273.9}{  2.4} \\
 0.825 & \res{ 1386.3}{ 70.3} & \res{ -527.7}{ 21.2} & \res{ 1179.4}{ 29.8} & \res{-2785.3}{ 30.9} & \res{   87.8}{  8.0} & \res{ -119.5}{  8.4} & \res{ -269.5}{  2.2} \\
 0.875 & \res{ 1636.2}{ 70.6} & \res{ -568.2}{ 21.3} & \res{ 1311.0}{ 31.0} & \res{-2996.6}{ 31.6} & \res{  109.3}{  8.7} & \res{ -165.5}{  9.1} & \res{ -256.2}{  2.0} \\
 0.925 & \res{ 1994.9}{111.6} & \res{ -674.3}{ 22.1} & \res{ 1553.8}{ 63.4} & \res{-3496.2}{ 38.8} & \res{  146.4}{ 15.4} & \res{ -261.0}{ 11.2} & \res{ -264.5}{  1.8} \\
 0.975 & \res{ 4032.8}{121.5} & \res{-1362.2}{ 25.5} & \res{ 3169.4}{ 67.0} & \res{-7444.4}{ 49.1} & \res{  329.2}{ 18.8} & \res{ -646.6}{ 17.3} & \res{ -473.7}{  2.1} \\
\end{tabular}                                               
\end{table}
\noindent

\end{document}